# Particle-in-cell simulations of laser beat-wave magnetization of dense plasmas

D. R. Welch, T. C. Genoni, C. Thoma, and D. V. Rose

Voss Scientific, LLC, Albuquerque, NM 87108

S. C. Hsu

Physics Division, Los Alamos National Laboratory, Los Alamos, NM 87545

## ABSTRACT

The interaction of two lasers with a difference frequency near that of the ambient plasma frequency produces beat waves that can resonantly accelerate thermal electrons. These beat waves can be used to drive electron current and thereby embed magnetic fields into the plasma [D. R. Welch et al., Phys. Rev. Lett. **109**, 225002 (2012)]. In this paper, we present two-dimensional particle-in-cell simulations of the beat-wave current-drive process over a wide range of angles between the injected lasers, laser intensities, and plasma densities. We discuss the application of this technique to the magnetization of dense plasmas, motivated in particular by the problem of forming high-$\beta$ plasma targets in a standoff manner for magneto-inertial fusion. The feasibility of a near-term experiment embedding magnetic fields using lasers with micron-scale wavelengths into a ~$10^{18}$ cm$^{-3}$-density plasma is assessed.

## I. Introduction

Recent research in magneto-inertial fusion (MIF)[1,2,3] has motivated a fresh look at methods to magnetize a plasma target by means other than in situ magnetic coils or the use of a compact toroid (CT) plasma as the target. The latter, by their very nature, limit the plasma $\beta$ to ~0.1–1, which in turn places constraints on the overall design of an MIF system. In the late 1970's, intense electron beams were used to form and magnetize plasmas[4] intended as a plasma target for the LINUS concept[5] involving liquid-liner implosions of a magnetized plasma target. In this work, we investigate the method of using laser-generated beat waves to resonantly accelerate thermal electrons in order to drive current and embed magnetic fields in a plasma.[6]

In the latter technique, known as beat-wave current drive,[7,8,9] two externally launched electromagnetic (EM) waves mix in a plasma medium, generating a plasma wave at their beat (difference) frequency. Increased understanding of beat-wave current drive and magnetic field seeding can impact numerous areas in magnetized high-energy-density laboratory physics (HEDLP), e.g., astrophysical plasmas,[10] MIF plasmas, high-power laser plasmas, and high-power plasma electronics. Recently, beat-wave fields from multiple overlapping lasers have been shown to produce stochastic ion heating for conditions relevant to the National Ignition Facility.[11] In a recent publication,[6] we demonstrated computationally how beat-wave current drive may be used to seed a magnetic field in an initially unmagnetized plasma. Using our state-of-the-art particle-in-cell (PIC) simulation capability, we further develop the scientific basis of this technique and apply it to MIF.[12,13,14,15,16,17] Although magnetic fields may be embedded in laboratory plasmas using both external coils[18] and laser-driven foils[19,20,21], beat-wave magnetization offers several potential advantages:

(1) standoff from the plasma is easily obtained via laser propagation;
(2) refraction of the injected high-frequency waves is negligible, and placement of the beat-wave interaction region within the plasma can be precise;
(3) beat waves are produced in a controllable direction depending on the angle $\theta$ between injected waves;
(4) current drive can be accomplished for thermal plasmas via control of the wave phase velocity which also depends on $\theta$; and
(5) topology of induced magnetic field structure can be controlled.

While early analyses of beat-wave current drive were essentially 1D,[8] more realistic modeling is now required to design actual high-power experiments on facilities such as OMEGA at the Laboratory for Laser Energetics (LLE)[22], the Z Accelerator at Sandia National Laboratories (SNL)[23], or the Trident Laser Facility at Los Alamos National Laboratory (LANL)[24]. For significant current drive, the beat-wave phase velocity ($v_{ph}$) and electron thermal velocity ($v_e$) must be comparable in order to accelerate a useful number of electrons. The electron distribution extends from a low-energy collisional component to a high-energy collisionless component requiring fully kinetic modeling. The modeling must also be at least 2D in order to handle essential experimental input parameters, such as the shapes, widths, and injection angle of the overlapping EM waves. Multi-dimensional modeling allows more accurate predictions of the spatial distribution of extended return currents and magnetic field and more realistic testing of predictions compared to 1D analysis.

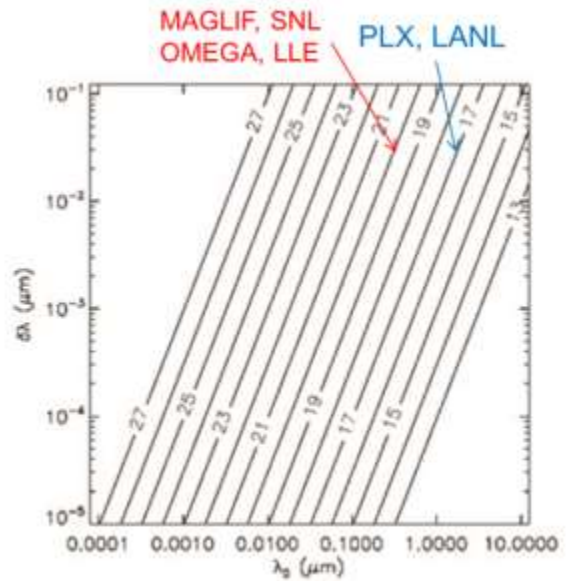

Figure 1. Contours of the logarithm of electron density (cm$^{-3}$) as a function of the center and difference wavelengths of the injected electromagnetic waves, satisfying $|\omega_1-\omega_2|=\omega_{pe}$, which is the beat-wave resonance condition. Indicated are plasma densities and laser conditions for some experiments.

The beat-wave current drive technique is applicable over a wide range of electron densities by tuning the injected waves according to the relationship $\omega_1 - \omega_2 \sim \omega_{pe}$, where $\omega_1$ and $\omega_2$ are the two injected wave frequencies and $\omega_{pe} \sim n_e^{1/2}$ is the electron plasma frequency. This technique has already been experimentally demonstrated using microwaves at very low densities.[9,25] Figure 1 shows the resonant electron density as a function of center ($\lambda_0$) and difference ($\delta\lambda$) wavelengths of the injected waves, indicating that infrared waves ($\lambda_0 \sim 1-10$ µm) to x-rays ($\lambda_0 \sim 10^{-4}-10^{-2}$ µm) can be used to potentially magnetize plasmas with electron density ranging from $10^{14}$ cm$^{-3}$ to $10^{27}$ cm$^{-3}$, respectively. It is worth noting that this range of densities covers everything from magnetic to inertial fusion. Assuming $\delta\lambda \sim 0.04$, beat-wave operating conditions for MagLIF at SNL[12], magnetized implosions on OMEGA at LLE,[16,17] and proposed magnetic field seeding of imploding spherical plasma liners on the Plasma Liner Experiment (PLX) at LANL[26], are also indicated in the figure. The efficiency of the current drive and thus the magnetic field generation process is of great importance for fusion energy applications.

Recent simulation results have begun to illustrate how to harness this technique to generate magnetic fields in an initially unmagnetized plasma.[6] Further simulations presented here explore the laser beat-wave interaction and the wave-particle coupling, and show how these interactions scale with plasma density, injected wave power and relative beam angle for designing near-term beat-wave current-drive experiments.

In this paper, we discuss the beat-wave current drive in two different regimes: (1) a $CO_2$ example with nominal 10-µm wavelength lasers interacting with $\sim 10^{16}$ cm$^{-3}$ plasma and (2) a 1-µm wavelength laser interacting with a $\sim 10^{18}$-cm$^{-3}$ plasma. We show that the basic scaling of the mechanism at the 10-µm laser wavelength can be applied to higher plasma densities and lower laser wavelengths, and present a proof-of-concept experimental scenario using existing laser technology. In Sec. II, we present the PIC simulation model used in the paper. We discuss the basic beat-wave and current-drive mechanisms in Sec. III. The physics limiting the extent and magnitude of the magnetic fields is discussed in Sec. IV. In Sec. V, we scale the technique to conditions relevant for HEDLP experiments and discuss the feasibility of a beat-wave current drive proof-of-concept experiment in a $\sim 10^{18}$ cm$^{-3}$-density plasma. Finally, we present conclusions in Sec. VI.

## II. Laser-plasma simulation model

For the simulations discussed in this paper, we used the LSP PIC code[27] which solves the relativistic Maxwell-Lorentz equations with inter-particle collisions. The code has the capability of injecting multiple propagating laser beams from all boundaries. The explicit LSP particle-advance algorithm (an implicit algorithm is also available) sums particle currents such that charge is conserved, i.e., the current density contribution from each particle is chosen such that it satisfies the continuity equation and, thus, Gauss's law. The algorithm is made energy conserving by scattering the individual particle current to, and gathering the electric field quantities for the momentum push from, the same staggered half-grid positions as the current.[28] The LSP energy-conserving algorithm does not require resolution of the Debye length to avoid numerical grid heating permitting orders of magnitude larger spatial extent for the beat-wave simulations. The LSP code models the particle scattering with complete generality by sampling each binary Coulombic interaction accounting for their individual probabilities.[29,30]

These simulations must resolve all relevant scale lengths required to model beat-wave generation and current drive. We use 20 grid cells per wavelength throughout the volume, sufficient to resolve the laser and beat-wave wavelengths. The electromagnetic Courant constraint for the explicit solution ensures that the laser period is temporally resolved as well. Because we are always well below the plasma critical density for the lasers, the plasma collisional skin depth ($c/\omega_{pe}$) and frequency are also well resolved. We have found that these spatial and temporal conditions yield numerically-converged results.

## III. Basic beat-wave current-drive mechanism

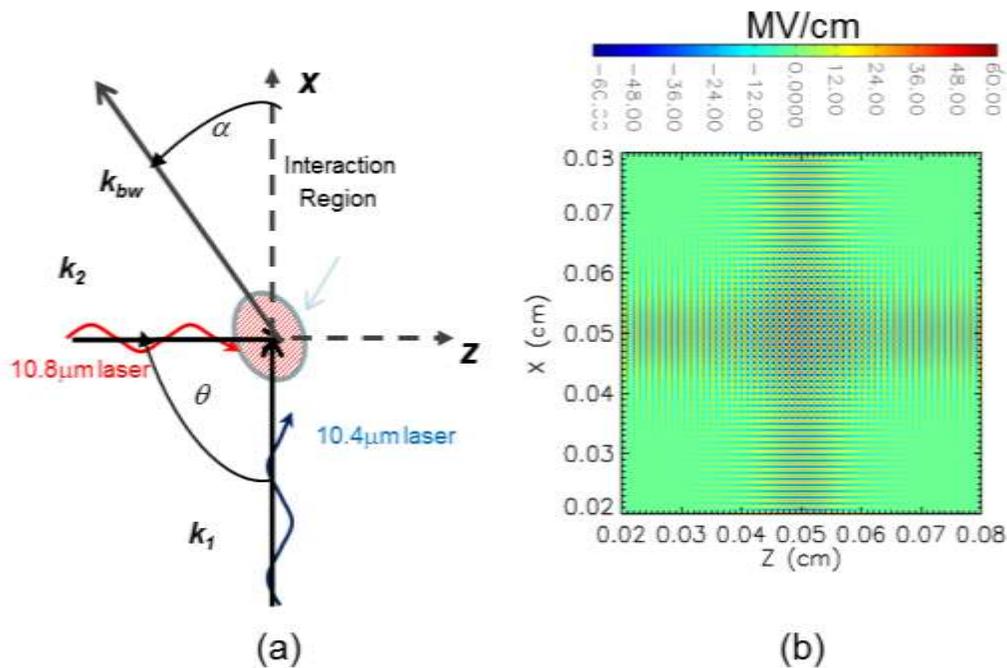

Figure 2. (a) In the 2D beat wave interaction, two waves intersect with angle $\theta$ in a plasma and create a beat wave at angle $\alpha$ with respect to the higher frequency $k_1$. (b) The superposition of two lasers with $\theta = 90°$ is shown.

The beat-wave current-drive method is based on launching electromagnetic waves into the plasma at well above the plasma's cut-off frequency, and relying on nonlinear mixing of the waves to generate a beat wave near the plasma frequency that is resonant with plasma electrons.[31,32,7] This resonance leads to wave-particle interaction and electron acceleration, which in turn leads to current drive and thus magnetic field generation. Consider a beat wave generated from the ponderomotive force ($\mathbf{E} \times \mathbf{B}$) of two intersecting injected waves with electric field $E$ polarization into the page and wave vector as exhibited in Figure 2. Since $\mathbf{k_{bw}} = \mathbf{k_1} - \mathbf{k_2}$, the angle between $\mathbf{k_{bw}}$ and $\mathbf{k_1}$ is $\alpha = \tan^{-1}(k_2 \sin\theta / [k_1 - k_2 \cos\theta])$ and the magnitude of the wave vector is

$$k_{bw} = \frac{k_1 - k_2 \cos\theta}{\cos\alpha}.$$

We see that the beat-wave phase velocity $v_{ph} = \omega_{pe}/k_{bw} \ll c$ for non-zero $\theta$. We confirmed in 2D simulations the theoretical scaling of the beat wave $v_{ph}$, $k_{bw}$, and direction with respect to the laser's conditions.[6]

Resonant interaction between the beat wave and the electron population can be exploited to accelerate electrons via Landau damping, and to drive current and generate magnetic fields. The wave-particle interaction will be most efficient when $v_{ph}$ is comparable to the electron thermal speed $v_e$. As a metric of this effect, we define $F = v_{ph}/v_e$. We have found that values of $F = 1.9$—2.7 are optimal.[6] For modest temperature plasmas ($\ll 1$ keV), current drive is most effective for $\theta > 45°$ where $v_{ph} \ll c$. The Landau damping mechanism accelerates a fraction of the electrons and is far more effective than high amplitude ($v_{ph} \sim c$) wave trapping of electrons for co-propagating lasers ($\theta = 0$). Two-dimensional simulations of 10-μm lasers interacting with a 10-eV, $10^{16}$-cm$^{-3}$ plasma have demonstrated current drive and that magnetic field production is an order of magnitude less when the lasers propagate with small $\theta$ compared with the opposing ($\theta=180°$) case.

## IV. Limiting physics for embedding magnetic field into the plasma

Although significant current drive in a non-relativistic plasma requires laser angles and $F$ in the proper range, the beat-wave-driven current is ultimately limited by the Alfven current. For a cylindrical beam, $I_A = 17000\gamma\beta$ A, where $\gamma$ and $\beta$ are the electron beam relativistic factor and speed relative to the speed of light, respectively. For the $10^{12-13}$ W/cm$^2$ CO$_2$ laser intensities considered here, the electrons are accelerated in the 500 kV/cm fields up to 1 keV or $\beta = 0.06$ with $I_A = 1$ kA. It should be noted that the Alfven current limitation is different for the 2D Cartesian simulations, involving current sheets, considered here. In this case, the current density limitation is roughly that of the cylindrical case, but the total current can be much larger depending on the transverse elongation of the current channel.

Given optimal values of $F$, higher intensities and higher beat-wave phase velocities can drive higher currents farther from the laser interaction region. The current drive extends in the direction of $k_{bw}$ away from the laser interaction region if the accelerated electrons have sufficiently small divergence. To demonstrate this effect in 2D LSP simulations with 10.4-μm and 10.8-μm wavelength lasers both with $3\times10^{12}$ W/cm$^2$ intensity at angle $\theta = 90°$, we set the initial plasma temperature at 10 and 50 eV in a 900-μm long, 400-μm wide, uniform density plasma of $2\times10^{16}$ cm$^{-3}$. For these two lasers that were roughly 4% different in frequency, the resonant plasma density $n_{res} = 1.4\times10^{16}$ cm$^{-3}$. We demonstrate later that there is a broad maximum in current drive above $n_{res}$. We find that the electron beam divergence is strongly dependent on temperature and roughly proportional to $1/F$. This observation leads to an intrinsic spatial limit on magnetic field penetration from the laser interaction region due to the spreading of the electron beam. This is a reasonable result when the beat-wave field supplies a purely longitudinal acceleration. The plasma temperature then provides the source transverse emittance for the beam. As seen in Figure 3, the electron beam divergence for the 10-eV plasma is roughly half that of the 50-eV plasma (0.17 versus 0.33 radians). Thus, to optimize the magnetic field volume for angles other than 180°, we desire laser configurations that reduce beam spreading. This technique permits the magnetization of plasmas well away from the interaction region and above $n_{res}$. Finally, simulations show that the transverse extent of the embedded magnetic field beyond that of the electron beam width is of order 1–2 collisionless plasma skin depths.

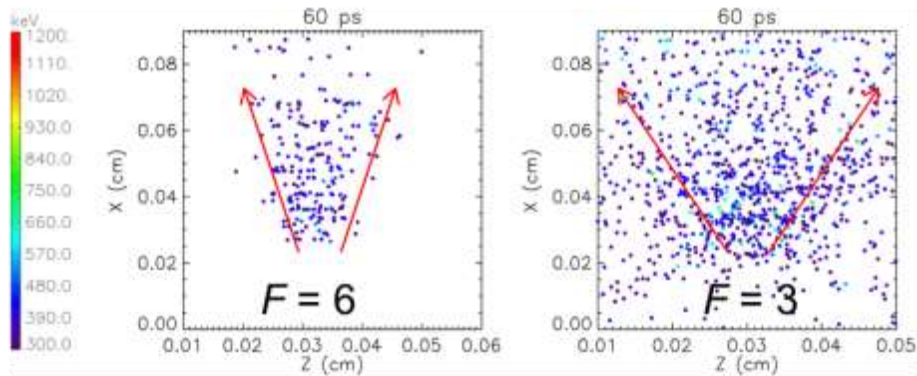

**Figure 3. For two lasers with $3\times10^{12}$ W/cm$^2$ intensities and $\theta = 90°$ injected into 10-eV ($F = 6$) and 50-eV ($F = 3$) plasma (uniform $2\times10^{16}$ cm$^{-3}$ density), high energy electrons (energy denoted by color) are plotted after 60 ps.**

For the MIF application, the embedded magnetic fields must persist as the plasma target is compressed. Propagating electrostatic and electromagnetic waves in magnetized plasma can introduce variations in particle and field quantities and have the potential to disrupt the current drive or cause more rapid decay of the magnetic fields. Several longitudinal modes propagating perpendicularly to the magnetic field are possible (E ∥ k ⊥ B). In regions of the plasma where density gradients are large, lower hybrid drift instabilities have been identified as a source of turbulence and anomalous resistivity.[33,34,35] Modeling this intermediate mode [$\omega \sim (\omega_{ce}\omega_{ci})^{1/2}$] requires resolution of the electron cyclotron radius. Lower frequency drift waves ($\omega < \omega_{ci}$) may be important in some regions of the plasma as well.[36] Finally, the Bernstein mode with frequency near $\omega_{ci}$ is also possible. The magnetized plasma can also support transverse EM waves propagating parallel to B (E $\perp$ k||B). Low frequency Alfven ($\omega \ll \omega_{ci}$) and intermediate whistler ($\omega_{ci} < \omega < \omega_{ce}$) waves are in this category. The effects of these waves on practical seeding of magnetic fields in the laboratory needs to be assessed in future work in that they will impact the lifetime of the magnetized channel after the current drive is removed.

## V. Application to high-density plasma experiments

We now consider the application of the beat-wave current-drive technique to HEDLP-relevant experiments. We must consider short-pulse micron-wavelength laser technology to meet the resonance condition in this regime. A reasonable and meaningful goal for such an experiment is to drive sufficient magnetic fields such that the cross-field energy transport is affected. For the embedded magnetic fields to limit thermal conduction across field lines, plasma electrons must be largely line tied with Hall parameter (defined as the ratio of the electron cyclotron to collision frequency) $H = 0.0011\ T\ (\text{keV})^{3/2}\ B(\text{T})/\rho\ (\text{g/cm}^3) \gg 1$.[12] For MIF experiments in which we desire significant fusion heating from DT alpha slowing, to inhibit alpha transport the alpha cyclotron radius $r_\alpha$ must be small compared with the fuel dimension $r_{fuel}$, i.e.,

$$r_\alpha = \frac{\gamma\beta c m_\alpha}{Z_\alpha e m_e B} < r_{fuel},$$

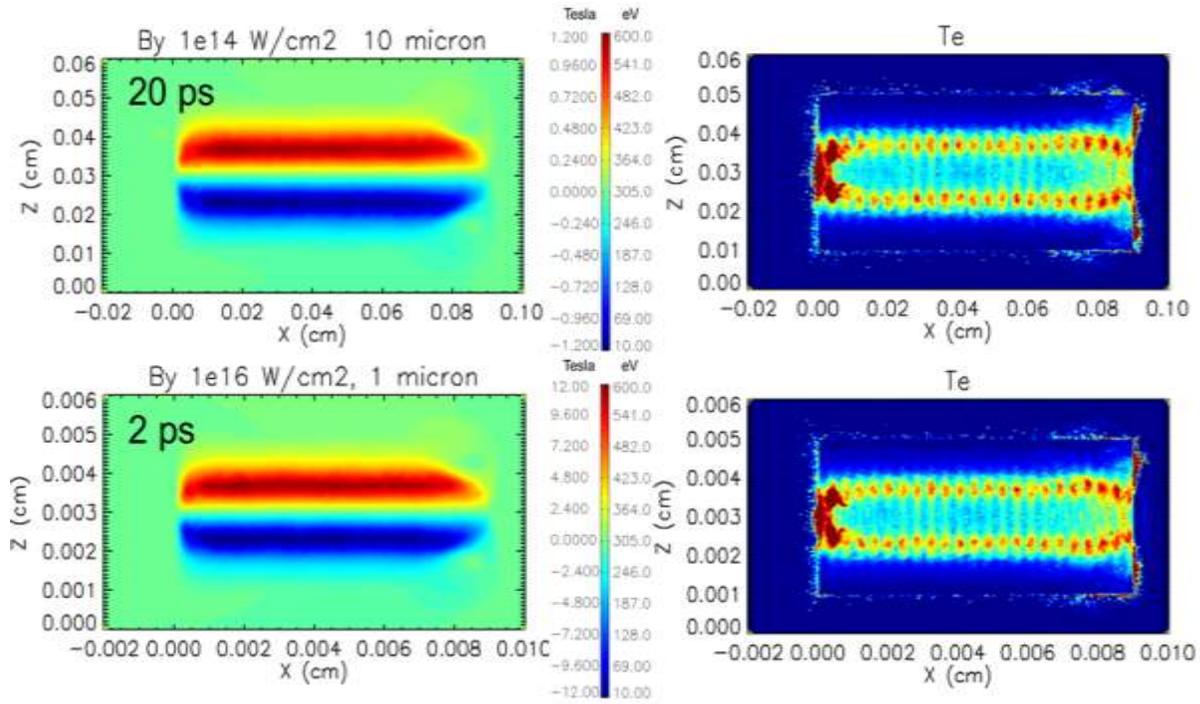

**Figure 4. For two scaled simulations, the magnetic field (left) and electron temperature (right) are plotted. The top row is scaled to a 10 μm laser wavelength at 20 ps, the bottom a 1 μm wavelength at 2 ps.**

where $\gamma\beta m_\alpha$ and $Z_\alpha$ are the alpha momentum and charge state. The required field to trap alphas scales as $B_\alpha = 4000$ T $(100$ μm $/r_{fuel})$. To attain these fields, we must rely on fuel compression after the beat-wave fields have been embedded. Inhibiting both thermal and alpha energy losses can have a positive impact on fuel density radius product requirements for fusion ignition and gain.

To demonstrate that beat waves can drive the appropriate fields, we consider two counter-propagating ($\theta = 180°$) lasers with a 4% variation in wavelength near 1 μm. This configuration scales nicely with the 10-μm laser simulation if we reduce all dimensions and time scales by 10 and all densities increase by 100. Similarly all EM fields increase by 10 and laser intensities by 100. Thus, the $n_{res}$ increases from $1.42\times10^{16}$ cm$^{-3}$ to $1.42\times10^{18}$ cm$^{-3}$. A quick test of this scaling uses a simulation setup with a 10-eV, uniform density, 400 (40) μm width and 900 (90) μm length plasma with opposing lasers with 4% difference frequency as shown in Figure 4. At the same scaled 20 (2) ps time, the $10^{14}$-W/cm$^2$ intensity, 10-μm wavelength laser drives nearly identical currents as well as the same plasma heating as a $10^{16}$ W/cm$^2$ intensity 1-μm lasers. The embedded magnetic fields are 10× higher for the higher intensity laser and similar in every detail. Peak fields are 1.2 (12) T and bulk heating of the plasma electrons to 200 eV. For this plasma density ($4.7\times10^{-6}$ g/cm$^3$) and assuming a temperature of 100 eV, we obtain $H \sim 5\ B(T)$. Embedded fields of order 1 T will affect thermal conduction. Thus, magnetic fields sufficient to affect plasma properties can be driven with the beat-wave technique over a large range of densities given the appropriate laser spot size.

## Transition from current drive to embedded magnetic field

We examine the details of the evolution of driving plasma currents to laser turn off and resulting embedded magnetic field for counter-propagating lasers with 1.04 and 1.08 µm wavelengths, 10 µm spot and $10^{15}$ W/cm$^2$ intensity in a $2\times10^{18}$-cm$^{-3}$ density 10-eV temperature plasma. The laser intensity reaches a peak at 1 ps and is flat until 10 ps and falls in 1 ps. Shown in Figure 5, this nominal laser-plasma configuration exhibits electron trapping in the beat wave fronts moving at $0.018c$ after $t = 0.5$ ps. At times 1—2 ps due to the electron velocity shear, the plasma return current interacts with the driven current, resulting in instability in the wings. The instability causes the return currents to become diffusive. By 5 ps, the plasma current channel has stabilized with current density as high as 100 MA/cm$^2$. The magnetic fields evolve inward from the wings as the instability progresses, as shown in Figure 6, producing a broader embedded field. By 5 ps, the peak field has reached 3 T.

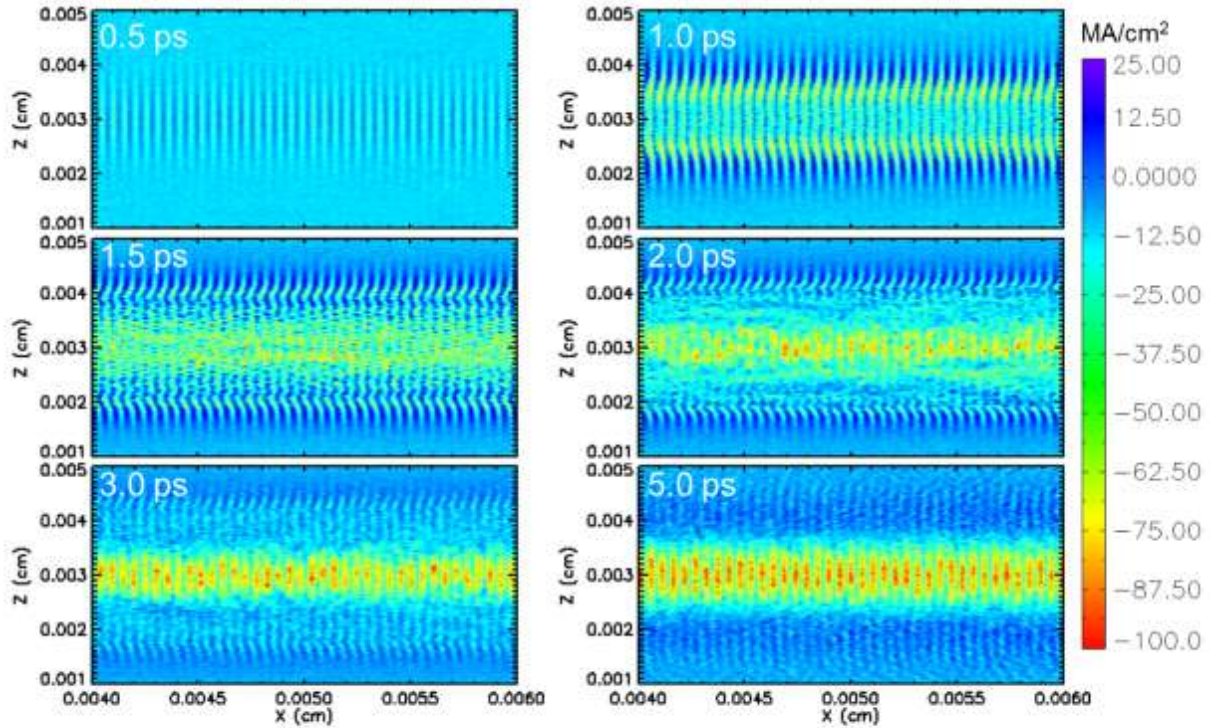

**Figure 5. For the nominal 1-µm laser-plasma configuration, the current density driven by opposing lasers is shown for times 0.5—5 ps. The plasma was initialized at 10 eV with $2\times10^{18}$ cm$^{-3}$ density.**

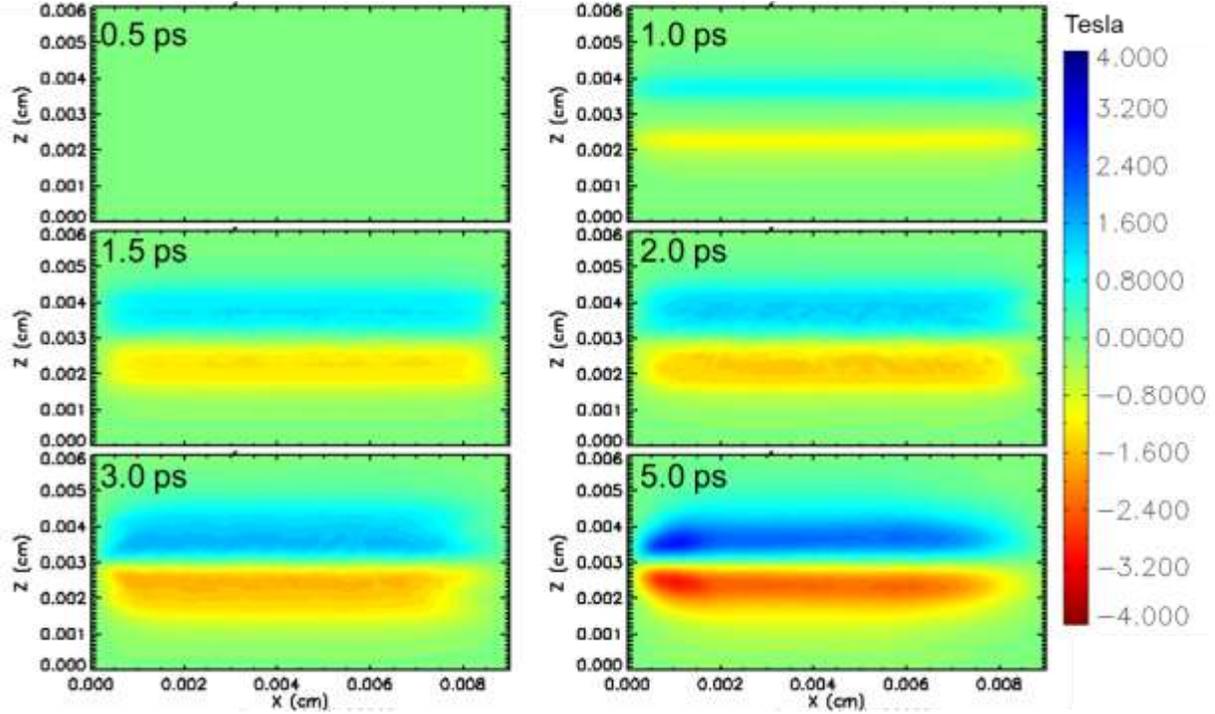

**Figure 6. The magnetic field is plotted for the nominal 1 µm laser-plasma configuration from 0.5—5 ps.**

The plasma current continues to be carried by the trapped plasma electrons with energies 100–500 eV. After 10 ps, we turn off the laser intensity in 1 ps. We can see the transition in Figure 7 from trapped electrons carrying the current to that of a homogenous plasma current by 12 ps. The corresponding magnetic fields have reached 4 T peak and have filled a large fraction of the plasma volume (see Figure 8). Despite the laser fields being off and the change in the character of the current carriers after 11 ps, the magnetic field strength does not fall. Thus, it is not merely a transient but an embedded field that is sustained by the long plasma current decay time,

$$\tau_m = \frac{4\pi\sigma L^2}{c^2},$$

where $\sigma$ and $L$ are the conductivity and scale length of the plasma with embedded field, respectively. For the 200-eV plasma temperature and .002-cm scale length of the magnetic field region simulated after 12 ps and assuming classical resistivity, we calculate $\tau_m$ = 10 ns which is considerably longer than the simulation time.

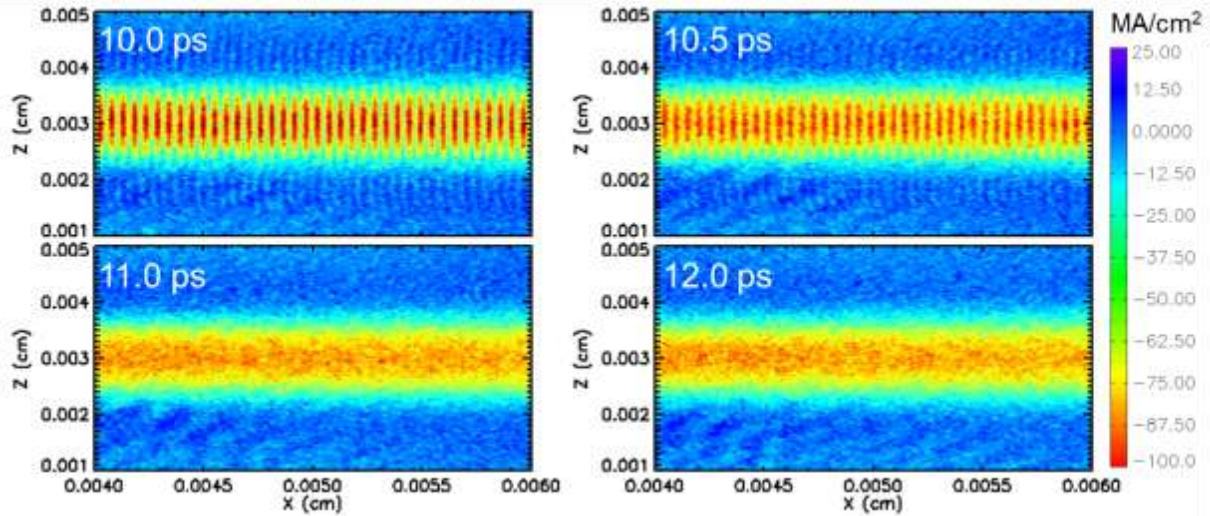

**Figure 7. The plasma current density is plotted as the lasers turn off from 10–12 ps. Note the transition from trapped electron current to homogenous plasma current.**

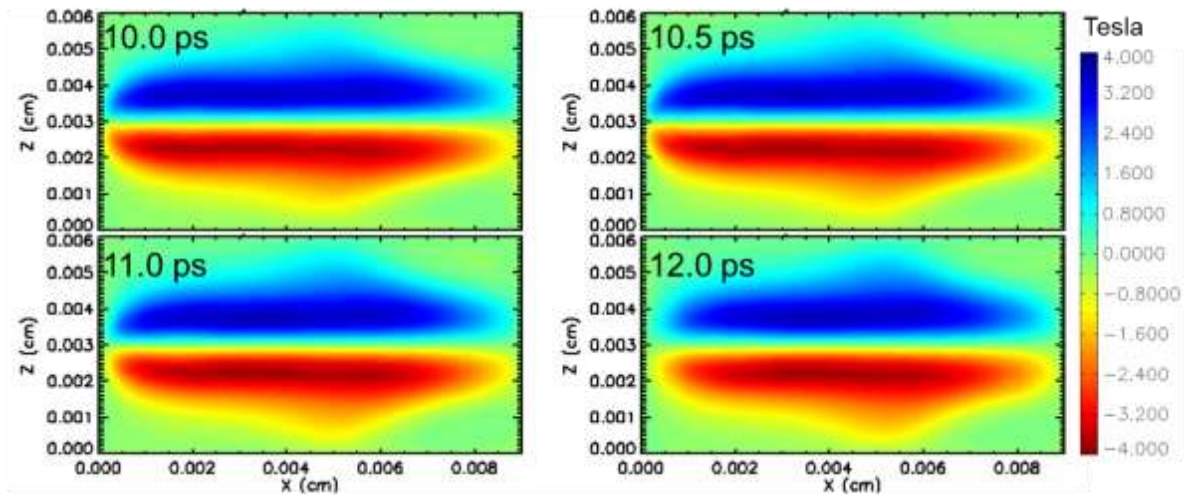

**Figure 8. The magnetic fields are shown for 10—12 ps as the lasers turn off.**

## Scaling of magnetic field with laser and plasma conditions

We have simulated several variations of the laser parameters including intensity ($10^{13}$—$10^{16}$ W/cm$^2$), spot size (2.5—15 µm), and plasma density (0.25—11 times the resonant plasma density of $1.4\times10^{18}$). The results are summarized in Figure 9. The peak magnetic field and the efficiency of injected laser energy to embedded magnetic field energy assuming 1-cm long plasma (because the laser range is much longer that the 0.009 cm plasma assumed) are plotted after 5 ps. Conditions are the nominal laser-plasma conditions unless specified in the plot. We see in (a) that the peak field and efficiency rise with laser intensities. The laser heating of the plasma is dependent on the inverse bremsstrahlung process and thus rises with laser intensity. For all these simulations near 1-μm laser wavelength, the plasma heats roughly in proportion to laser intensity with the fraction of laser energy going into heating $f_{therm} = 0.025 n/n_{res}$. As the heating progresses, a larger fraction of plasma electrons can thus be trapped between beat wave fronts. This nonlinear effect enhances the scaling of magnetic field with laser intensity to almost linear. Peak efficiency approaches 0.2% at $10^{16}$ W/cm$^2$ intensity.

Keeping the laser power constant (3.5 J/cm laser energy in 5 ps) but varying the spot size in Figure 9(b), we find an optimal spot size for both peak field and efficiency from 5—10 μm which is roughly 1.5—3 plasma skin depths. We also find an optimal plasma density relative to the resonant density of 4—6 as seen in (c). At the optimal density $6n_{res}$, peak magnetic field is nearly 4 Tesla and efficiency 0.25%. The field production does not fall off rapidly at larger densities indicating a broad density acceptance. On the low density side, the field production does fall off to 0.3 T by for $0.25n_{res}$.

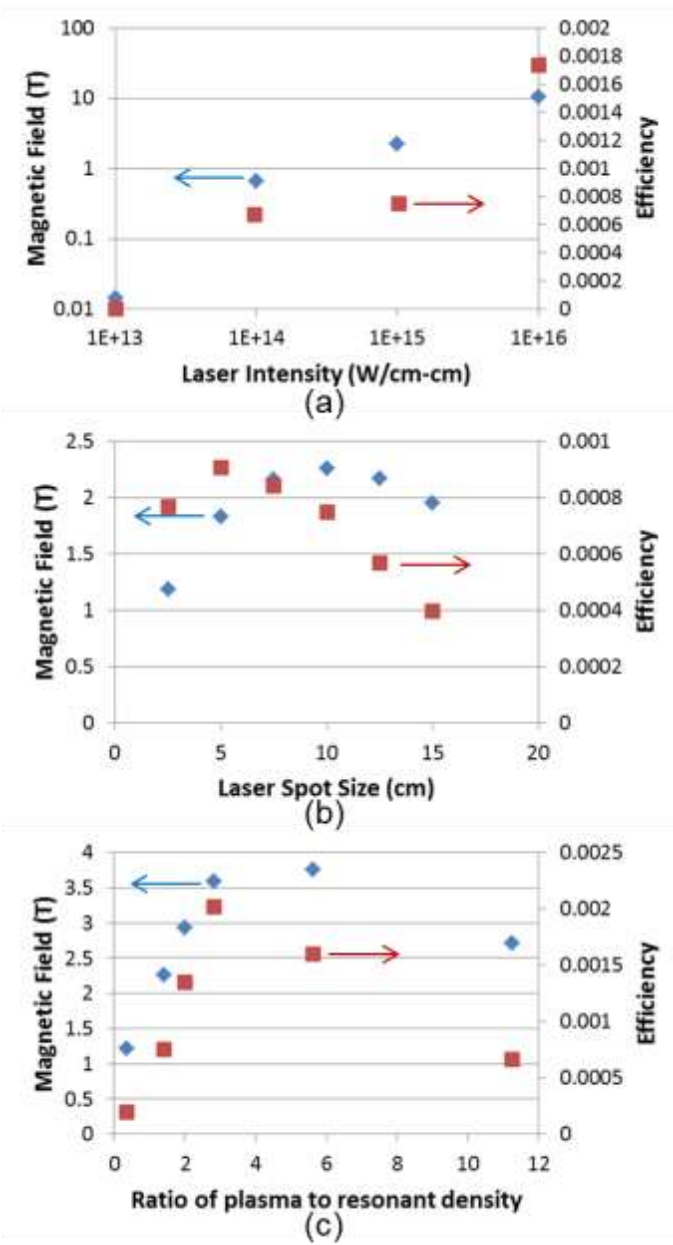

**Figure 9.** The peak magnetic fields and energy efficiency are plotted for (a) varying laser intensity, (b) laser spot size, and (c) ratio plasma density to resonant density.

## V.A Application to MagLIF and OMEGA experiments

We now discuss the use of lasers with difference frequencies applicable to MagLIF experiments at SNL's Z facility[12] and magnetized implosions on the OMEGA laser at LLE.[16] In both Z and OMEGA experiments, the $> 10^{20}$ cm$^{-3}$ initial fuel densities require higher frequency lasers with wavelengths in the range 250—500 nm. This corresponds to critical densities of order $10^{22}$ cm$^{-3}$. Experiments on OMEGA have used pulsed magnets to initially seed 5—16 T fields in 3 atmospheres of D$_2$ gas (roughly $10^{20}$ cm$^{-3}$ atomic density or $\rho$ = 0.0003 g/cm$^3$) within a 430-µm radius, 1.5-mm long CH cylinder.[37] Assuming $T$ = 100 eV gives $H \sim 1$ initially yielding modest effect on thermal conduction. Compression of the plasma increases $H$ further because of the effect of both field compression and plasma heating. The OMEGA laser drove a high velocity implosion trapping the field to 3000—3600 T with an inferred compression ratio of order 20. Although large, these fields are only 1/5 those required to trap alphas as discussed earlier. The seed field allowed higher temperatures at stagnation. Slutz, *et al.*,[12] have proposed experiments on the Z machine that involve magnetized (> 10 T), pre-heated (100—500 eV), 0.5-cm radius and long cylindrical liner filled with DT fuel also giving $H \sim 1$. In the envisioned experiment, the pre-heat is supplied by the Z-Beamlet laser with 5—10 kJ of energy at 500 nm into an initial 0.5-cm fuel radius. Given a compression ratio of 20, the resulting magnetic field is sufficient to produce alpha cyclotron radius order that of the compressed fuel. Calculated fusion gains as high as 1000 have been predicted.[38]

To achieve the required current drive at $10^{20}$ cm$^{-3}$ plasma density, we need to reduce scale sizes down a factor of 2—4 to the appropriate 250—500 nm laser wavelength which taking into account a 10$n_{res}$ limit on density will permit 6x10$^{19}$—1.3x10$^{20}$ cm$^{-3}$ initial densities. The scaling also predicts proportionately higher fields from 8—16 T. The laser intensities increase to 4—9x10$^{15}$ W/cm$^2$ with laser spot size decreasing to 2.5—5 µm. The laser spot should be highly elliptical as modeled in the 2D simulations in order to drive these fields in the largest plasma volume. Although the energy of each laser channel (3.5 J/cm) is held constant in the scaling, the number of channels (order 10—100 for 100—1000 µm scale targets) must increase to fill the required volume. The total amount of laser energy is order 0.4—4 kJ. This could be accomplished by using many lasers or sweeping the elliptical spot in a given direction. In this section, we have shown in laser-plasma simulations that the required magnetic fields of > 10 T from the beat-wave interaction in densities approaching that for both MAGLIF and OMEGA experiments are feasible; however, the laser technology is not "off the shelf."

## V.B Simulations of a proof-of-concept experiment on the Trident Laser Facility

We now examine the details of the evolution of the beat-wave driven plasma currents and magnetic fields for counter-propagating lasers with 1.054 and 1.064 µm wavelengths and 10-µm spot size. Although not ideal, with a reduced difference frequency compared to previous calculations (1% versus 4%), these laser wavelengths are achievable with current technologies and are being considered for experiments on the Trident Laser Facility[39] at LANL. In our calculations, the intensity for the 1.054-µm laser is fixed at 10$^{15}$ W/cm$^2$ intensity with the second 1.064-µm laser intensity varying from 10$^{12}$—3×10$^{14}$ W/cm$^2$. For both, the laser intensity reaches a peak at 1 ps and is flat through 5 ps. The resonant beat wave plasma density for this laser configuration is $n_{res}$ = 8.9×10$^{16}$ cm$^{-3}$ with wave phase velocity $v_{ph}$ = 0.0047$c$. For electron thermal velocity $v_{the}$, the ideal plasma temperature at which $F = 1.9$—$2.7$ corresponds to a 1—3 eV plasma electron temperature range. Plasma heating beyond this range is not advantageous.

In 2D simulations, the lasers propagate into a uniform density (1—$8\times10^{17}$-cm$^{-3}$) 1—3-eV temperature plasma. For $3\times10^{13}$ W/cm$^2$ second laser intensity and a 3eV, $2\times10^{17}$-cm$^{-3}$ plasma (referred to as nominal), strong beat waves are excited with 1000 kV/cm magnitude and persist for the 5-ps long simulation. The electrons are trapped in the beat wave fronts moving at $0.0047c$ reaching nearly 5-MA/cm$^2$ current density. The qualitative behavior of the current drive is similar to that seen in Figure 5. The plasma electron density and $<v_x>$ show that a large percentage of the electrons are being carried in the beat waves in conveyer belt fashion. The density is modulated over 30% with mean velocity approaching that of phase velocity ($<v_x> \sim ¾ v_{ph}$). Accelerated electron energies are as high as 300 eV.

The beat waves drive magnetic fields up to 1.4 kG, somewhat smaller than calculated for the lasers with larger difference frequency. These fields are quite smooth indicating a uniform current density of nearly 15 µm width. This width is typically the summation of width of the laser spot and 1—2 skin depths. There is heating of the plasma electrons past 20 µm width with peak 17 eV temperature, well above the optimum value of $F$.

We have simulated several variations of the laser parameters including the second laser intensity ($10^{12}$—$3\times10^{14}$ W/cm$^2$) as well as the plasma density from 2—9 $n_{res}$. Although fields are somewhat weaker, results are qualitatively similar to those presented in Figure 9. The variation of magnetic field with second laser intensity ($n = 2n_{res}$) is shown in Figure 10. The peak magnetic field increases roughly linearly with this intensity. The field also increases somewhat with density from 0.21 T at n = $2n_{res}$ reaching 0.35 T for density $n = 9n_{res}$ at which density the field has been shown to reach a maximum from previous simulation scans. Direct measurement of these sub-Tesla fields will be a key challenge for such an experiment.

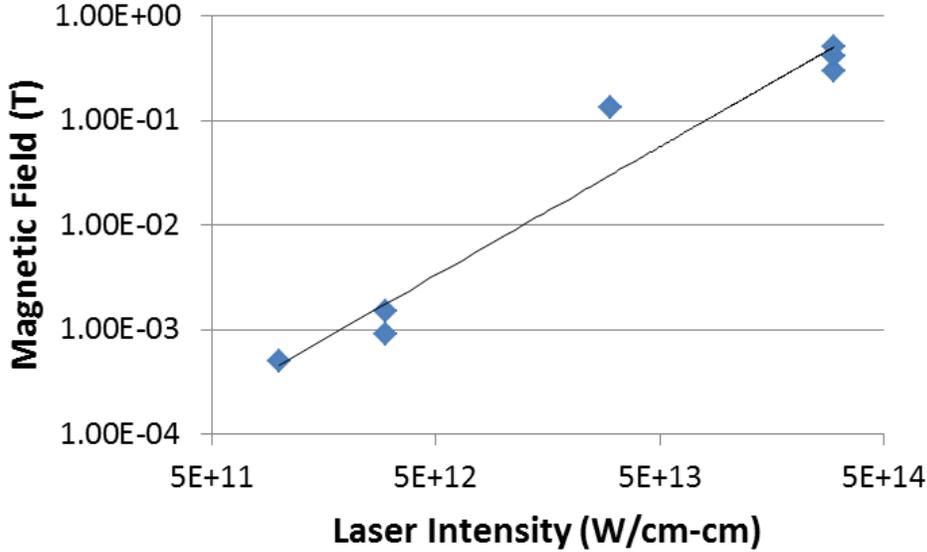

**Figure 10.** The peak magnetic field is plotted after 5 ps for varying laser intensity of the 1.064 μm laser. The 1.054 μm laser intensity was held fixed at $10^{15}$ W/cm$^2$ and $n = 2n_{res}$.

## VI. Conclusions

In this paper, we have described in detail the scientific basis for standoff magnetization of an unmagnetized plasma via beat-wave current drive. Specifically, we explored via 2D PIC simulations the scaling of beat-wave production, current drive, and magnetic field generation, and also applications experiments with plasma densities as high as $10^{19}$ cm$^{-3}$ using lasers with ~1-μm wavelength. The limitations of volumetric magnetization, in particular the Alfven current limit and electron beam divergence have also been described. By decreasing the laser wavelength and maintaining 1—5% difference between the lasers, the beat wave technique can be used at the higher densities required for magnetization of dense plasmas. For plasma densities in the range $10^{18-19}$ cm$^{-3}$, a 1-μm wavelength is sufficient with relatively short pulse or 1—10 ps. We have shown that for reasonable laser energies (10 J), a long channel (1 cm), 40 μm width channel can be created with up to 0.2% efficiency and produce > 10 T field. This field is sufficient to magnetize present and proposed experiments on Z and OMEGA, although the required volume of plasma to be magnetized demands multiple or spatially swept laser channels. The merging process of multiple magnetized plasma filaments is a topic for future work. This can be accomplished remotely eliminating the need for intrusive magnet coils. For these plasma densities approaching $10^{20}$ cm$^{-3}$, laser wavelengths order ≤0.5 μm are required.

A small-scale experiment to investigate and demonstrate this technique is the next logical step. Using available technology for laser wavelengths, we have shown that reasonably strong magnetic fields can be generated with currently available 1.054 μm and 1.064 μm wavelength lasers. The scaling of embedded magnetic field with laser/plasma parameters suggested in this paper can be verified in a proposed experiment at the Trident Laser Facility at LANL. In addition to *producing* significant magnetic fields, the *lifetime* of the fields in the embedded plasma relating to the magnetic decay time in the presence of possible anomalous resistivity needs to be assessed.

## Acknowledgments

We acknowledge useful discussions with D. Q. Hwang and Y. C. F. Thio and excellent code support from R. E. Clark. This work was supported by the Office of Fusion Energy Sciences of the U. S. Dept. of Energy under contract nos. DE-AC52-06NA25396, DE-FG02-05ER54835, and DE-S0010698.